# Experimental Evidence of Ferroelectric Negative Capacitance in Nanoscale Heterostructures

Asif Islam Khan,[1] Debanjan Bhowmik,[1] Pu Yu,[2] Sung Joo Kim,[4] Xiaoqing Pan,[4] Ramamoorthy Ramesh,[2,3] and Sayeef Salahuddin[1]

[1] Electrical Engineering and Computer Sciences, University of California Berkeley, CA – 94720,

[2] Physics, University of California, Berkeley, CA 94720

[3] Material Science and Engineering, University of California, Berkeley, CA 94720

[4] Materials Science and Engineering, University of Michigan, Ann Arbor, MI 48109

Corresponding Author: Sayeef Salahuddin (sayeef@eecs.berkeley.edu)

## ABSTRACT

We report a proof-of-concept demonstration of negative capacitance effect in a nanoscale ferroelectric-dielectric heterostructure. In a bilayer of ferroelectric, $Pb(Zr_{0.2}Ti_{0.8})O_3$ and dielectric, $SrTiO_3$, the composite capacitance was observed to be larger than the constituent $SrTiO_3$ capacitance, indicating an effective negative capacitance of the constituent $Pb(Zr_{0.2}Ti_{0.8})O_3$ layer. Temperature is shown to be an effective tuning parameter for the ferroelectric negative capacitance and the degree of capacitance enhancement in the heterostructure. Landau's mean field theory based calculations show qualitative agreement with observed effects. This work underpins the possibility that by replacing gate oxides by ferroelectrics in MOSFETs, the sub threshold slope can be lowered below the classical limit (60 mV/decade).

CMOS scaling is facing a fundamental barrier stemming from the Boltzmann statistics that dictates that a minimum voltage must be applied to effect an-order-of-magnitude increase in the current[1,2]. This means that CMOS voltage and transistor power dissipation cannot be downscaled arbitrarily. Therefore, it has been suggested that without introducing fundamentally new physics in transistor operation, an end to scaling is inevitable[3]. In that pursuit, it was proposed[4] that the minimum voltage requirement could be overcome if the ordinary gate oxide could be replaced by another stack that provides an effective negative capacitance. The key to overcoming the Boltzmann limit by negative capacitance lies in the fact that in a series combination of a negative and a positive capacitor, the total capacitance becomes larger than its constituent positive capacitor. Notably this is just the opposite of what happens in a classical series combination of two positive capacitors, where the total capacitance is always reduced. For a MOSFET, a negative gate capacitance can make the total capacitance, looking into the gate, larger than the semiconductor capacitance. Then to induce the same amount of charge in the channel, one would require a smaller voltage than what would be required classically. This, in turn, means that the gate voltage could be reduced below the classical limit. In order to understand how an effective negative differential capacitance could appear in a ferroelectric material, we start by noting that capacitance, $C$ can be related to the energy of the element, $U$ by,

$$C = \left[ d^2U / dQ^2 \right]^{-1} \qquad (1)$$

where, $Q$ is the charge stored in the capacitor. For a nonlinear capacitor, the energy can be expanded as

$$U = \alpha P^2 + \beta P^2 + \gamma P^2 \qquad (2)$$

where, $\alpha, \beta, \gamma$ are material dependent constants. The coefficient $\alpha$ is negative for a ferroelectric capacitor[5]. The physics of this negative $\alpha$ is well described within the Landau's mean field based theory of ferroelectrics[5]. Leveraging on this fact, it can be shown by using eq. (1) and (2) that ferroelectric capacitance is indeed negative in a certain range of $P$, around $P = 0$. In Fig. 1(b), the negative capacitance region of the ferroelectric energy landscape (curve 1) is shown inside the dotted rectangular box. However, ferroelectric negative capacitance cannot be directly measured or accessed in an isolated ferroelectric because the negative capacitance state is an unstable one which results in the hysteretic jumps in the polarization (*P*) vs. electric field *(E)* characteristics.

Properties of a ferroelectric material can be strongly modulated by the temperature. Within the phenomenological Landau model, this dependence is captured by the fact that $\alpha = \alpha_0(T-T_c)$ is strongly temperature dependent and $\alpha$ is negative only up to the Curie temperature, $T_c$. Based on this physics, the negative capacitance and degree of capacitance enhancement in FE-DE heterostructure can be tuned by changing the temperature. Figure 1(c) shows the simulated capacitance of a FE-DE heterostructure as a function of temperature and compares it to that of the constituent DE. Also shown in Fig. 1(c) is the FE capacitance in the heterostructure as well as the voltage amplification factor[25] at the FE-DE interface, $(1+C_{DE}/C_{FE})^{-1}$. Notably, only at a certain temperature range, the FE capacitance is stabilized in the NC region and capacitance enhancement occurs in that temperature range, where FE capacitance is negative and the amplification factor is greater than one. The temperature dependence is explained in details in the Supplementary Information Section. Figure 1(d) compares the simulated *C-V* characteristics of a $Pb(Zr_{0.2}Ti_{0.8})O_3$ (PZT)-$SrTiO_3$ (STO) heterostructure with that of the constituent STO at different temperatures.

To test this hypothesis, we fabricated FE-DE bilayer capacitors using Pb(Zr$_{0.2}$Ti$_{0.8}$)O$_3$ as the ferroelectric and SrTiO$_3$ as the dielectric material. Details of the synthetic approaches, device fabrication and measurements are described in the Supplementary Information.

We first focus on the capacitance of a 28 nm PZT - 48 nm STO bi-layer capacitor. Fig. 2(a) shows the *C-V* characteristics at 100 kHz of a PZT (28 nm)-STO (48 nm) bilayer capacitor at different temperatures and compares it to the capacitance of a 48 nm STO. As predicted from the simulation, the capacitance of PZT-STO sample is larger than that of the STO capacitor at elevated temperatures. The evolution of *C-V* curves of the PZT-STO and the STO capacitors with temperature has very similar trends as compared to those obtained by simulation, shown in Fig. 1(d). Fig. 2(b) shows the capacitance of the PZT-STO bi-layer and an isolated dielectric STO as a function of temperature. Also shown in Fig. 2(b) are the extracted capacitance of the PZT in the bilayer[6] and the calculated voltage amplification factor at the PZT-STO interface. At roughly 225°C, the bi-layer capacitance exceeds the STO capacitance. *This means that beyond this temperature, the capacitance of the 76 nm thick bi-layer (STO: 48nm+PZT:28 nm) becomes larger than that of 48 nm STO itself.* Fig. 2(c) shows the capacitance as a function of frequency at *T* = 300°C. We see that enhancement in capacitance is retained even at 1 MHz thereby indicating that defect mediated processes are minimal, if any and therefore the enhanced capacitance cannot be attributed to such effects. The measured effective dielectric constant of the bi-layer and the isolated STO as well as simulated STO dielectric constant[18] is shown in Fig. S4(b)[Supplementary Information]. Despite the fact that, at room temperature, the measured $\varepsilon_r$ for the STO layer is smaller than the simulated STO $\varepsilon_r$ as well as the highest reported $\varepsilon_r$ in thin film STO[8], at high temperatures, our measured STO $\varepsilon_r$ is as high as that obtained from simulations. The bilayer capacitance enhancement is observed at elevated temperatures, where the measured STO dielectric constant ($\varepsilon_r$) is as high as the theoretical limit. This fact precludes the possibility that the bilayer capacitance is unduly compared to an STO thin film that has lower $\varepsilon_r$ due to undesired 'dead layer' at the Au-STO interface that is otherwise absent in the bilayer structures. This is discussed in details in the supplementary information section.

Fig. 3 shows the capacitance and permittivity of 3 different samples (sample no 2-4) of PZT-STO bi-layers of different thicknesses, an isolated STO (sample 1) and an isolated PZT sample (sample 5). All the bi-layer samples show an enhancement in overall permittivity and capacitance as the temperature is increased.

As far as electrostatic boundary conditions are concerned, FE-DE superlattices are similar to FE-DE bilayer heterostructures. Similar enhancement in dielectric constants has indeed been observed in $BaTiO_3/SrTiO_3$[9] and $PbTiO_3/SrTiO_3$ superlattices[10,11]. However, Maxwell-Wagner effect was suggested as a possible origin of the enhanced dielectric constant in the super lattices.[8] Catalan et al.[8] showed that the MW mediated dielectric enhancement depends on the number of interfaces in the superlattice and the enhancement essentially dies out by 10 kHz. Hence, it is unlikely that in our heterostructures, which have just one FE-DE interface,[23] MW effects can cause dielectric enhancement at frequencies as high as 1 MHz. By knowing the individual permittivities of the two layers and measuring the leakage component of the bi-layer capacitor, the overall permittivity that could be obtained due to MW effect can be estimated.[19] These estimated values do not correspond to the measured value for our samples (see Supplementary Information Section, Fig. S5). Therefore, we conclude that the enhancement in permittivity in our samples is unlikely to have come from leakage mediated effects. Furthermore, existence of domains in ferroelectric superlattices has been confirmed in Ref. 11 via X-ray diffraction spectroscopy. However, domain wall movement with applied bias in the ferroelectric alone cannot explain capacitance enhancement of the superlattice, simply because of the fact that without negative capacitance contribution from the ferroelectric, the superlattice capacitance cannot be larger than the constituent STO capacitance.[11]

Negative capacitance is a direct consequence of the negative curvature of the energy landscape in the ferroelectric material. A number of other physical systems also have negative terms in their energy profile. One example is the exchange correlation between two closely spaced 2D electron gases (2DEG). Experimentally, a negative compressibility was measured between two closely spaced 2DEG in a modulation doped GaAs/AlGaAs heterostructure at cryogenic temperature[12]. Very recently, enhanced capacitance was measured in epitaxial LAO/STO heterostructure[13], also at cryogenic temperature, and was explained in terms of a negative capacitance that could arise due to similar exchange correlation.[13,14] The advantage of ferroelectric material based systems comes from the fact that the negative energy terms are reasonably large at room temperature thereby removing the need for cryogenic operation.

Very recently a demonstration of <60 mV/decade has been achieved in a FET by putting a polymer ferroelectric on the gate.[20] In addition, multiple theoretical analysis[21,22] have investigated the design space of a negative capacitance FET. Our demonstration in a simple capacitor structure and in a different material system (crystalline oxides as opposed to polymer FE) is expected to confirm the overall concept of negative capacitance and also provide new insights into the effect.

To summarize, we have experimentally demonstrated that in a ferroelectric-dielectric series capacitance network, the total capacitance, contrary to what happens in a classical series combination, could be larger than the constituent dielectric capacitance. While integration of ferroelectrics on silicon is a major technology challenge, SrTiO$_3$ can be used as a perovskite template for growing FE on silicon.[17] In that case, the FE/STO/semiconductor structure has to be engineered by tuning the FE thickness so that negative capacitance effect reaches STO/semiconductor interface. Going beyond overcoming the "Boltzmann limit" of transistor power dissipation, the enhancement in capacitance can be very useful for ultra-dense dynamic random access memory (DRAM) applications. Increased capacitance beyond the classical limit can also lead to novel super capacitor structures. Low Curie temperature ferroelectric materials could make it possible to obtain the enhancement at room temperature.

This work was supported in part by FCRP Center for Materials, Structures and Devices(MSD) and also by Office of Naval Research (ONR). The authors acknowledge Rehan Kapadia and Ali Javey for their help with gold evaporation.

25. Voltage amplification ratio at the FE-DE interface is the ratio of the voltage at the across the DE to the total voltage applied across the FE-DE.

**FIGURE CAPTIONS**

**Figure 1:** (a) Schematic of the experimental stack. $SrTiO_3$ (STO) and $Pb(Zr_{0.2}Ti_{0.8})O_3$ (PZT) form the bi-layer. Au and SrRuO3 (SRO) are used as top and bottom contacts respectively for capacitance measurements. (b) Energy landscapes of a dielectric and a ferroelectric capacitor are characterized by single well (curve 1) and double well (curve 2) shapes respectively. The negative capacitance region of the ferroelectric energy landscape (curve 1) is shown inside the dotted rectangular box. The energy landscape of the series combination of the two capacitors is obtained by adding the energies of the component capacitors (curve 3). The negative curvature of the ferroelectric makes the curvature of the FE-DE bi-layer smaller than that of the dielectric material. Hence, the bilayer capacitance is larger than the constituent DE capacitance. (c) Capacitance of a FE-DE bi-layer capacitor as a function of temperature. The FE here is a 40 nm $Pb(Zr_{0.2}Ti_{0.8})O_3$ and the DE is a 10 nm dielectric with $\varepsilon_r = 200$. Also shown in this figure is the capacitance of the constituent DE and FE in the heterostructure as well as the voltage amplification factor at the FE-DE interface. In a certain temperature range, the FE is stabilized in the negative capacitance state when the bi-layer capacitance is larger than the constituent DE capacitance. (d) Calculated *C-V* characteristics of a $SrTiO_3$ capacitor and a $Pb(Zr_{0.2}Ti_{0.8})O_3$ - $SrTiO_3$ bi-layer capacitor at $T=T_A$, $T_B$ and $T_C$. The thickness ratio of PZT and STO in the heterostructure is 4:1. The capacitance of this heterostructure is plotted as function of temperature in Supplementary Fig. S3(a). The Landau coefficients of PZT and STO are taken from ref. 18 for calculation.

**Figure 2:** (a) Comparison of *C-V* characteristics of a PZT (28 nm)-STO (48 nm) and an STO (48 nm) sample at different temperatures. The *C-V* characteristics are qualitatively similar to those obtained using simulation, shown in Fig. 1(d). PZT-STO capacitance is higher than the STO capacitance at elevated temperature. (b) Capacitances of the samples at the symmetry point as functions of temperature measured at 100 kHz and the extracted PZT capacitance in the bilayer. Symmetry point refers to the cross point of the C-V curves obtained during upward and downward voltage sweeps. Capacitance enhancement occurs above 225 $^{o}$C at 100 kHz. Also shows here is the calculated amplification factor at the FE-DE interface. (c) Capacitances of the samples as functions of frequency at 300 $^{0}$C.

**Figure 3:** Comparison of dielectric constant (a) and capacitance (b) of several PZT-STO samples with those of STO and PZT at 100 kHz at different temperatures. In (b), the capacitance of the constituent STO in each of the bilayers is shown by small horizontal line.

# Experimental Evidence of Ferroelectric Negative Capacitance in Nanoscale Heterostructures

Asif Islam Khan,[1] Debanjan Bhowmik,[1] Pu Yu,[2] Sung Joo Kim,[4] Xiaoqing Pan,[4] Ramamoorthy Ramesh,[2,3] and Sayeef Salahuddin[1]

[1] Electrical Engineering and Computer Sciences, University of California Berkeley, CA – 94720,

[2] Physics, University of California, Berkeley, CA 94720

[3] Material Science and Engineering, University of California, Berkeley, CA 94720

[4] Materials Science and Engineering, University of Michigan, Ann Arbor, MI 48109

Corresponding Author: Sayeef Salahuddin (sayeef@eecs.berkeley.edu)

## SUPPORTING ONLINE MATERIAL

**Simulation of Ferroelectric-Dielectric Heterostructure Capacitors:**

According to the Landau theory[1], the energy landscape of a ferroelectric (FE) and a (non-linear) dielectric (DE) can be expressed as an even polynomial of the polarization, *P*. For a series combination of a FE and a DE capacitor with $l_f$ and $l_d$ as the FE and DE layer thicknesses, the total energy of the combination can be written as,

$$U_{f+p} = l_f (\alpha_f P_f^2 + \beta_f P_f^4 + \gamma_f P_f^6 - E_f \cdot P_f) + l_d (\alpha_d P_d^2 + \beta_d P_d^4 + \gamma_d P_d^6 - E_d \cdot P_d)$$

(S1)

where $\alpha_i$, $\beta_i$ and $\gamma_i$ with $i \equiv f$ or $d$ are the anisotropy constants of the FE and DE material respectively, $P_f$ and $P_d$ are the polarization in the FE and DE capacitors respectively and $E_f$ and $E_d$ are the electric fields inside the FE and DE capacitors respectively. To satisfy the Kirchhoff's voltage law in the system,

$$V = E_f l_f + E_d l_d \qquad (S2)$$

and applying Gauss law at the interface,

$$\varepsilon_0 E_d + P_d = \varepsilon_0 E_f + P_f \qquad (S3)$$

Combining Eq. 3, 4 and 5, the total energy of the system can be rewritten as,

$$U_{f+d} = l_f(\alpha_f P_f^2 + \beta_f P_f^4 + \gamma_f P_f^6) + l_p(\alpha_d P_d^2 + \beta_d P_d^4 + \gamma_d P_d^6) - V\frac{P_f l_f + P_d l_d}{l_f + l_d} + \frac{l_f l_d (P_f - P_d)^2}{\varepsilon_0 (l_f + l_d)}$$

(S4)

For a given applied voltage, $V$, the combined free energy $U_{f+d}$ as in (S4) is minimized with respect to $P_f$ and $P_d$ under the constraints of eq. (S2) and (S3).

The charge between the top and the bottom electrode is:

$$Q = \varepsilon_0 E_d + P_d = \varepsilon_0 E_f + P_f \qquad (S5)$$

The capacitance is then calculated using the relation $C = \frac{dQ}{dV}$.

The values of PZT and STO Landau coefficients are taken from ref. 2.

Voltage amplification ratio at the FE-DE interface, which is defined as the ratio of the voltage at the FE-DE interface to the total voltage applied across the FE-DE bilayer can be calculated as $(1+C_{DE}/C_{FE})^{-1}$.

**Stabilization of Ferroelectric Negative Capacitance:**

Eq. (S4) can be further simplified to get to an intuitive picture of ferroelectric negative capacitance stabilization. The last term in Eq. (S4) is the electrostatic energy arising due to the polarization mismatch at the interface. It has been shown in a number of references that in a FE-PE system, this polarization mismatch is costly in energy and the system would adopt a nearly uniform polarization throughout all the layers. Thus, with $P_f = P_d = P$, Eq. (S4) can be rewritten as

$$U_{f+d} = l_f(\alpha_f P^2 + \beta_f P^4 + \gamma_f P^6) + l_p(\alpha_d P^2 + \beta_d P^4 + \gamma_d P^6) - P\frac{V}{l_f + l_d} \quad (S6)$$

With no applied voltage, the total energy of a series combination of a FE and a PE capacitor is thus an addition of the energy of the individual capacitors. Fig. 1(a) shows the individual and combined energy landscapes in a series combination of a FE and a PE capacitor at V=0. For the system depicted in Fig. 1(a), the FE and PE energy landscapes are matched so that the combined energy landscape minima corresponds to the negative capacitance region of the FE landscape and hence the FE capacitor is stabilized in the negative capacitance state. In this system, the equivalent capacitance of series combination will be larger than the capacitance of the component PE capacitor.

**Effect of Temperature on the Ferroelectric-Dielectric Heterostructures:**

Ferroelectric and dielectric properties are sensitive to temperature and hence for given FE-DE system of with a specific thickness combination, temperature can be used as a tuning parameter to match of the energy landscape of the FE and DE capacitors. Fig. S2(a) shows the simulated capacitance of a 4:1 PZT-STO heterostructure as a function of temperature and compares it to the capacitances of an isolated STO and an isolated PZT of the same thicknesses as in the heterostructure. Fig. 1(d) (main text) shows a comparison of the simulated *C-V* characteristics of the STO capacitor and the PZT-STO capacitor at $T = T_A$, $T_B$ and $T_C$. At $T = T_A$, PZT-STO shows hysteresis in *C-V* characteristics and PZT-STO capacitance is smaller than of the STO capacitor around *V= 0*. At $T=T_B$, the PZT-STO hysteresis has decreased and equivalent capacitance is larger than that of the STO capacitor around zero bias. Finally at $T=T_C$, the PZT-STO hysteresis has collapsed. In order to explain the temperature dependence, we show the energy landscapes of the PZT-STO heterostructure for two different temperatures, $T_A$ and $T_B$ in Fig. S2(b) and S2(c). In a ferroelectric, as the temperature increases, the double wells in the energy landscape come closer to each other and also become shallower and flatter making the ferroelectric capacitance increase. At the Curie temperature, the double wells develop into a single minimum, the capacitance peaks and a FE-to-DE phase transition occurs for the ferroelectric. Beyond Curie temperature, the energy landscape in the DE phase becomes steeper again around the single minima making the capacitance decrease with temperature. As shown in fig. S2(a), both FE and FE-DE have qualitatively the same shape of capacitance vs.*T* characteristics with the apparent Curie temperature of the FE-DE system shifted to a lower temperature. It is because the energy landscape of the series combination essentially has the same shape as that of a ferroelectric one and due to the energy contribution from the dielectric capacitor, the double wells are shallower and closer than those of the FE system. As shown in fig. S2(a), at $T=T_A$, the minima of series combination falls outside the negative capacitance region of the FE landscape and hence capacitance enhancement is not observed. As temperature increases from $T_A$ to $T_B$, the minima of the series combination moves closer to the origin faster than that of the component ferroelectric capacitor due to contribution of the dielectric energy contribution and hence beyond a critical temperature, the minima of series combination correspond to the negative capacitance region of the ferroelectric landscape resulting in the capacitance enhancement of the series combination. Above the Curie temperature of the FE, $\alpha$ of the ferroelectric becomes positive and hence FE capacitor behaves like an ordinary positive capacitor. Hence the Curie temperature of the ferroelectric is the upper limit of the negative capacitance temperature range in a FE-DE heterostructure.

**Growth, Device Fabrication and Characterization:**

PZT is a robust room temperature ferroelectric with a remnant polarization of ~ 80 μC/cm$^2$ [3]. PZT has a tetragonal crystal structure with unstrained *c*- and *a*-axis lattice parameters of 0.4148 and 0.3953 nm and $c/a \approx 1.05$[4]. Unstrained STO has a cubic crystal structure with $c = a = 0.3905$ nm[5] and remains paraelectric down to 0 K.[8] SRO has a pesudocubic crystal structure with lattice parameter of 0.393 nm. Well matched lattice structures for all these materials allow coherent and epitaxial growth of atomically smooth PZT/STO/SRO heterostructures on STO (001) substrates.

Three different types of heterostructures were grown using pulsed laser deposition technique: STO/SRO, PZT/SRO and PZT/STO/SRO. Stoichiometric PZT, STO and SRO targets were ablated at a laser fluence of ~ 1 J cm$^{-2}$ and a repetition rate of 15, 5 and 15 for PZT, STO and SRO respectively. During growth, the substrate was held at 720 $^0$C for SRO and STO and at 630 $^0$C for PZT. Lower growth temperature of PZT is adopted to prevent the evaporation of volatile Pb and also to reduce dislocation density at the PZT/STO interface due to small lattice mismatch at lower temperature[6]. PZT and SRO were grown in an oxygen environment at 100 mTorr and STO in 250 mTorr. After growth, the heterostructures were slowly cooled down to room temperature at 1 atm of oxygen at a rate of ~ 5 $^0$C min$^{-1}$ to optimize oxidation. SRO thickness in all the samples is ~ 30 nm. Surface topography, X-ray diffraction spectrum and TEM cross-sectional images of representative PZT-STO samples are shown in Fig. S3.

Gold top electrodes were ex-situ deposited by e-beam evaporation, and then patterned using standard lithographic technique into circular dots of 16 – 60 µm diameter. Capacitance measurements were performed using an HP 4194A Impedance/Gain-Phase Analyzer.

**Calibration of STO Dielectric Constant at Elevated Temperatures:**

Our measured STO dielectric constant at room temperature (~200) is lower than the highest reported STO (50 nm) permittivity (~ 300),[8] which raises the possibility that there exists a dead layer at the Au-STO interface that may reduce the intrinsic dielectric constant of STO and that is absent in Au-PZT interface in the bilayers. In this regard, it is to be noted that bilayer capacitance enhancement is not observed at high temperatures only and to the best of our knowledge, temperature dependence of $\varepsilon_r$ of single crystal STO thin films (50-100 nm) on SRO buffered STO (001) substrate at higher than 300 K has not been reported in literature. Hence, in order to benchmark our measurements at high temperatures against standard values, we simulated STO dielectric constant using Landau theory.[2] The simulated STO $\varepsilon_r$ is in excellent agreement with the measured $\varepsilon_r$ of STO at the reported temperature range in Ref. 8 (see Fig. S4(a)). Fig S4(b) shows the temperature dependence of STO simulated using Landau theory and compares that with our measured dielectric constants of STO and PZT-STO bilayer. It is important to note that simulated and measured permittivities are in the same range. Hence, at elevated temperatures, the dielectric enhancement of PZT-STO bilayer is large enough to yield capacitance enhancement, even if we compare the measured PZT-STO bilayer capacitance with the simulated STO capacitance.

**Analysis of Maxwell-Wagner Effect in Bilayer Systems**

In order to model Maxwell-Wagner effect in a bilayer of two insulators, the leakage conductivity of each of insulating layer is modeled as a resistor parallel to the layer capacitances (Fig. S5(a))[7]. There are two limiting cases that one could consider. If the leakage across PZT is extremely high so that $R_{PZT} = 0$, then the capacitance of the circuit will be equal to that of STO. In this limit, of course, the capacitance can only be as high as that of STO. The other limiting case is when due to extreme leakage, one could imagine $R_{STO} = 0$. In that case, the total capacitance will be equal to that of PZT and if, and only if PZT capacitance is larger than STO, one would see a total capacitance that is larger than STO. For our case, this second condition is the one that needs to be investigated to make sure that the observed capacitance enhancement is not a result of leakage mechanisms.

Now, for given values of capacitances and resistances of PZT and STO layers, the equivalent admittance of the PZT-STO bi-layer (as shown in fig. S5(a)) can be calculated using the following equation,

$$Y = \frac{1}{R_{PZT-STO}} + i\omega C_{PZT-STO} = \left[ (\frac{1}{R_{PZT}} + i\omega C_{PZT})^{-1} + (\frac{1}{R_{STO}} + i\omega C_{STO})^{-1} \right]^{-1} \qquad (S7)$$

where, $R_{PZT-STO}$ and $C_{PZT-STO}$ are equivalent resistance and capacitance of the PZT-STO combination, $\omega$ is the angular frequency and $i = \sqrt{-1}$. To make a fair comparison, we take the measured PZT-STO resistance, $R_{PZT-STO}$ as the PZT resistance, $R_{PZT}$ ($R_{PZT} = R_{PZT-STO}$). This makes sense in the limit when $R_{STO}=0$. $C_{PZT}$ and $C_{STO}$ are set equal to the measured capacitance of isolated PZT and STO sample. The values are tabulated in table S1. The STO resistance, $R_{STO}$ was varied by six orders of magnitude around the isolated STO resistance, $R_{STO-base}$ and the capacitances of the bilayer stack for different values of STO resistances were calculated using eq. (S7). Figure S5(b) shows the ratio of the calculated PZT-STO capacitance to the measured PZT-STO capacitance, $C_{PZT-STO,calculated}/C_{PZT-STO,measured}$ as a function of $R_{STO}/R_{STO,base}$ at 300 and 400 $^0$C. It is observed that at 300 $^0$C, M-W effect cannot give the degree of enhancement in PZT-STO sample as observed for any value of $R_{STO}$. At 400 $^0$C, the STO resistance has to be smaller than that of the isolated STO layer by more than four orders of magnitude to get the same enhancement. In figure S5(c) we show the ratio of the calculated PZT-STO admittance angle (from Eq. (S7)) to the measured PZT-STO admittance angle, $\theta_{PZT-STO,calculated}/\theta_{PZT-STO,measured}$ as a function of $R_{STO}/R_{STO,base}$ at 300 and 400 $^0$C. It can be readily understood by comparing figures S5(b) and S5(c) that at 400$^0$C, the resistance of STO that would result in the same degree of enhancement of bilayer capacitance as in the actual sample would yield a higher admittance angle than what is obtained in the actual samples. On the other hand, the values of $R_{STO}/R_{STO,base}$ that gives the same admittance angle as the measured value show that $R_{STO}$ cannot be less than a few percent of $R_{STO,base}$. From Fig. S5(b), in this range of $R_{STO,}$ the capacitance calculated from MW effect would much lower than the measured value. Hence, we believe that leakage alone cannot give rise to the enhancement of capacitance to the extent as seen in these samples.

**SUPPLEMENTARY FIGURE CAPTIONS**

**Figure S1:** A ferroelectric-dielectric bilayer capacitor under an applied voltage, $V$.

**Figure S2:** (a) Simulated capacitance of a PZT-STO bi-layer capacitor with PZT and STO thicknesses of $t_{PZT}$ and $t_{STO}$ respectively, an isolated STO with a thickness of $t_{STO}$ and an isolated PZT with a thickness of $t_{PZT}$ as a function of temperature. $t_{PZT}:t_{STO} = 4:1$. (b,c) Energy landscapes of the series combination at two different temperatures, $T_A$ (b) and $T_B$ (c).

**Figure S3:** Structural characterization of the heterostructures. (a) AFM topography image of typical PZT-STO sample surfaces showing 0.5 nm RMS roughness. (b) XRD θ-2θ scans around (002) reflections of a PZT (42 nm)-STO (28 nm)-SRO (30 nm) and a PZT (39 nm)-SRO (30 nm) sample. The STO (002) peak is buried within the STO (002) substrate peak. The diffraction pattern confirms the c-axis orientation of the PZT films *without* any contribution from the a-axis oriented domains and impurity phases. (c) Cross-sectional HRTEM images of different interfaces of a PZT-STO sample. PZT/STO film on SRO-buffered STO have in-plane epitaxy to the substrate with atomically sharp interfaces.

**Figure S4:** (a) Comparison of STO dielectric constant simulated using Landau model[2] with measure dielectric constant of 50 nm STO reported in Ref. 8. (b,c) Comparison of dielectric constant (b) and capacitance (c) of a PZT (28 nm)-STO (48 nm) sample with those of STO samples at 100 kHz as functions of temperature. $r=1$ line refers to the required bilayer dielectric constant to achieve $C_{PZT-STO} = C_{STO}$ at a certain temperature.

**Figure S5:** (a) Maxwell-Wagner circuit representation of an insulating bilayer system[1]. (b, c) Ratio of calculated and measured PZT-STO capacitances (b) and admittance angles (c) as functions of the STO resistance in the bilayer at 300 $^0$C and 400 $^0$C.

**SUPPLEMENTARY TABLE CAPTION**

Table S1: Measured capacitance and resistance values of isolated PZT, isolated STO and PZT-STO bilayer capacitors at 100 kHz at 300 and 400 $^0$C.

**TABLE**

Table S1:

| Quantity | 300 $^0$C | 400 $^0$C |
|---|---|---|
| $R_{PZT-STO}$ | 150 kΩ | 130 kΩ |
| $C_{PZT-STO}$ | 63 pF | 72 pF |
| $C_{PZT}$ | 53 pF | 71 pF |
| $R_{STO, base}$ | 500 kΩ | 250 kΩ |
| $C_{STO}$ | 48 pF | 44 pF |

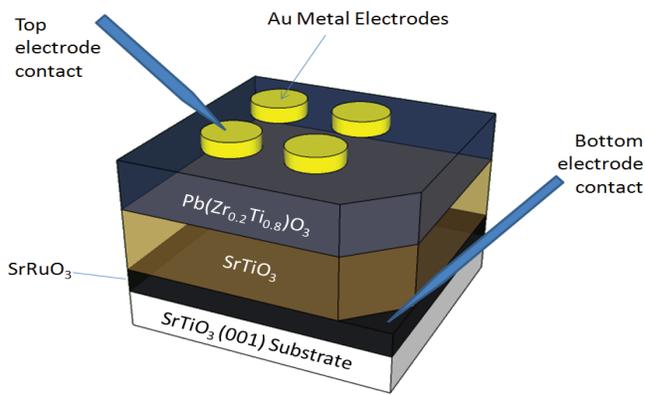
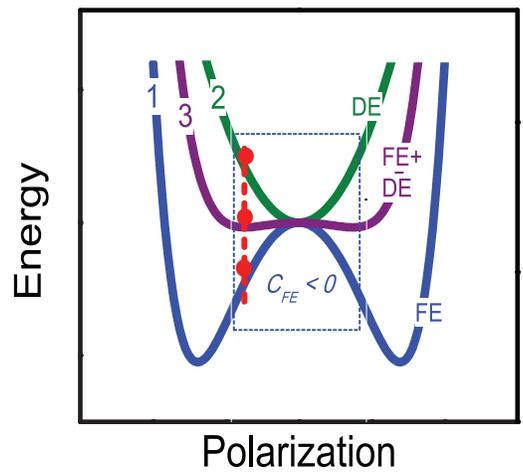
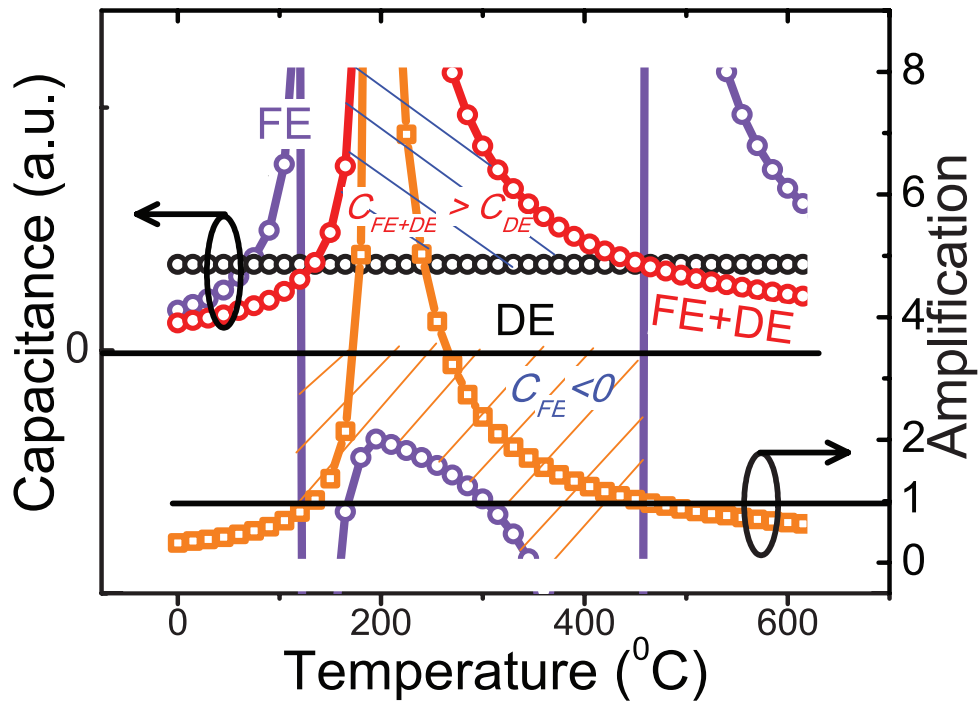
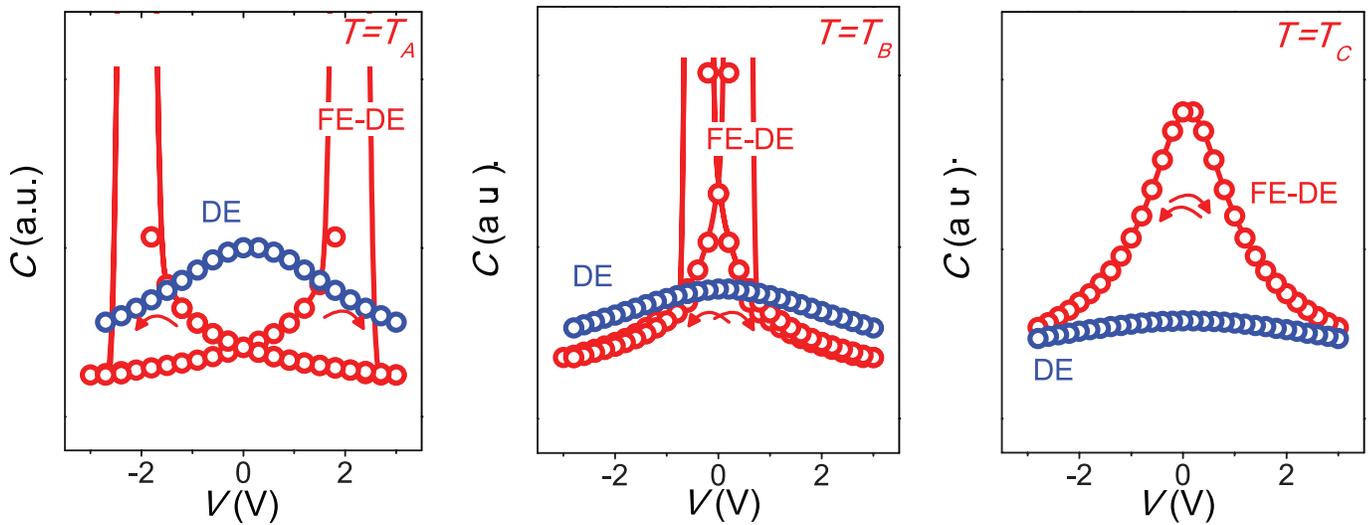

**FIGURE 1**

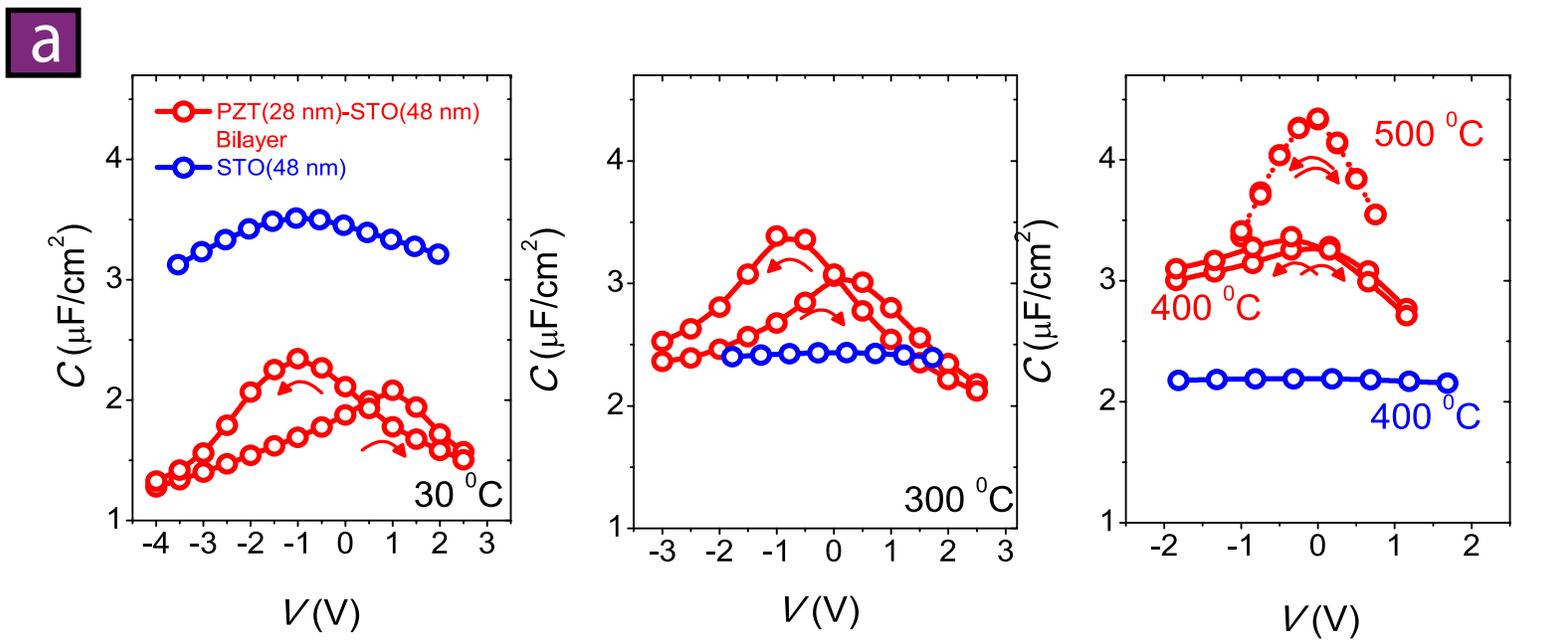
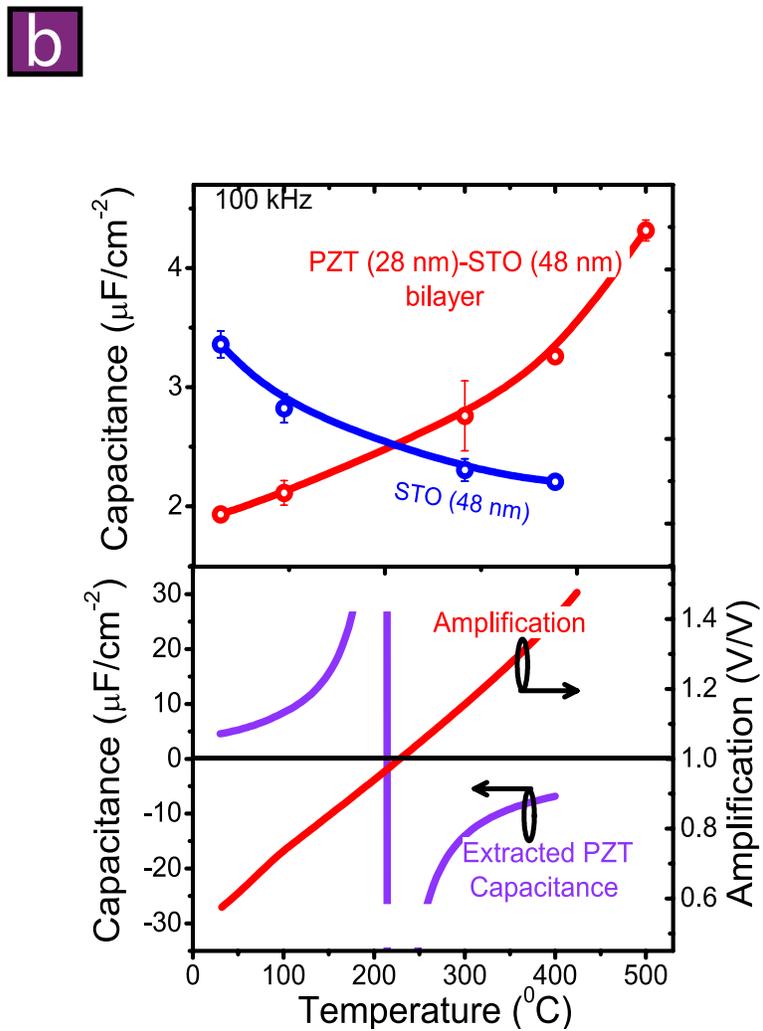
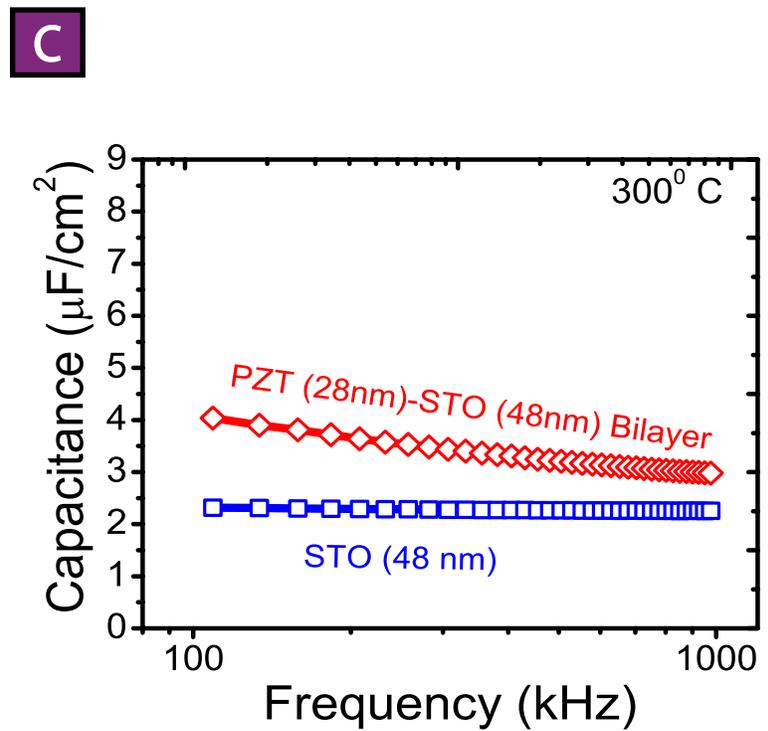

**FIGURE 2**

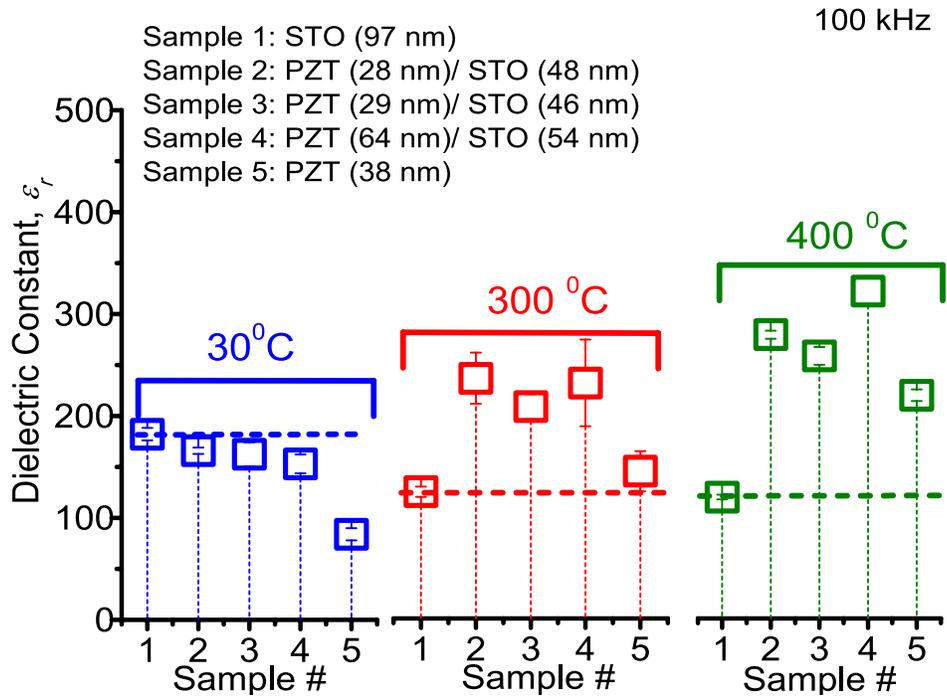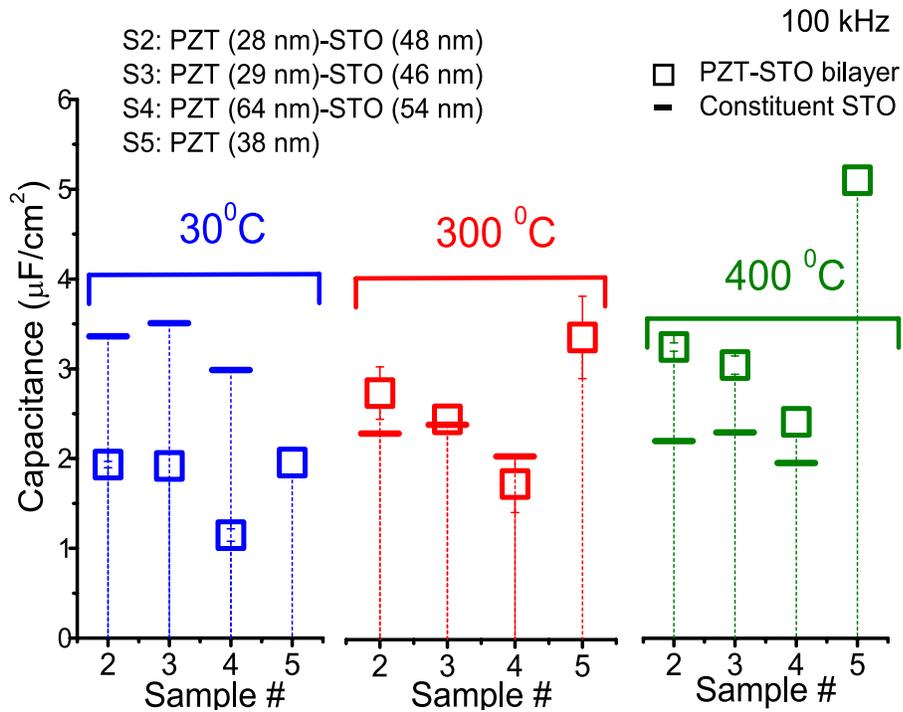

FIGURE 3

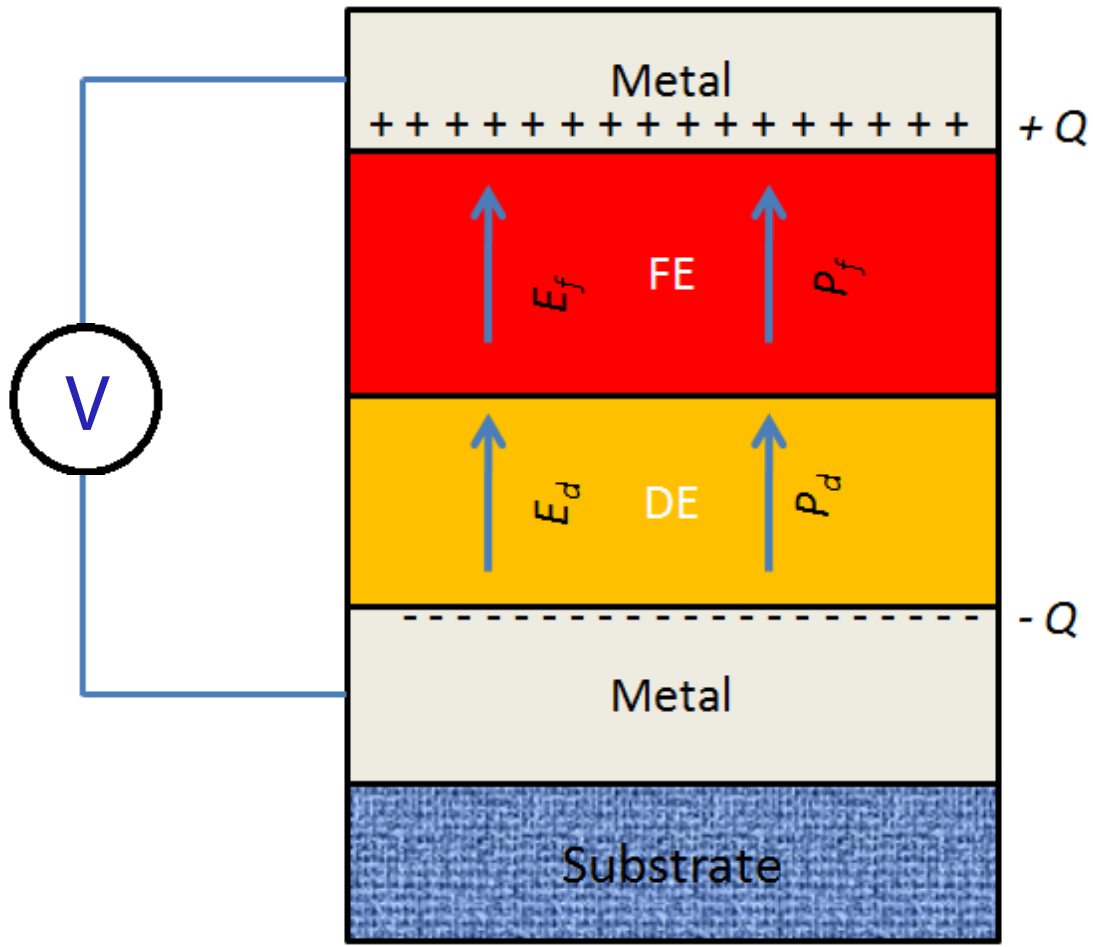

FIGURE S1

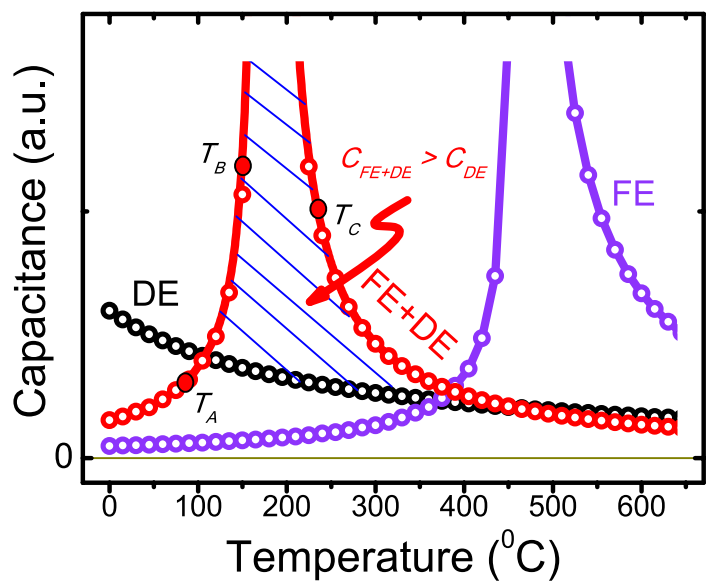
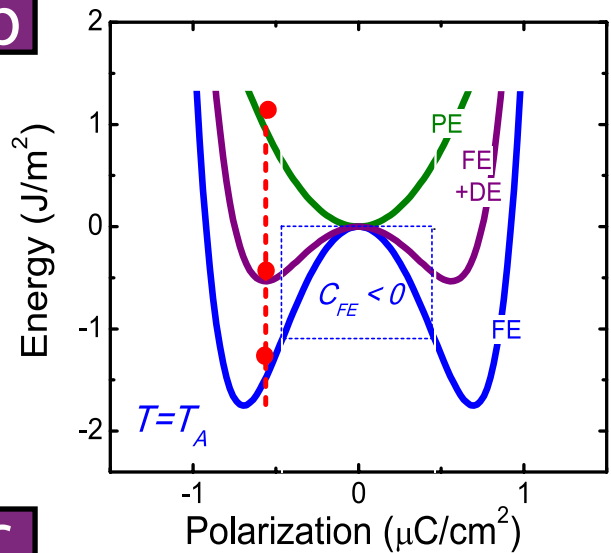
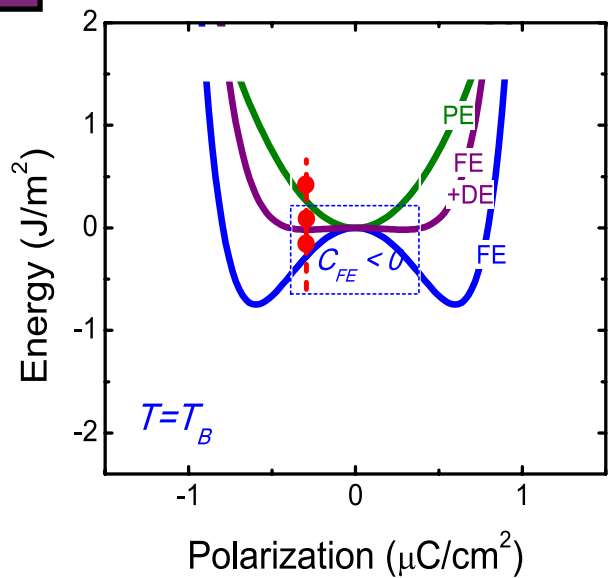

**FIGURE S2**

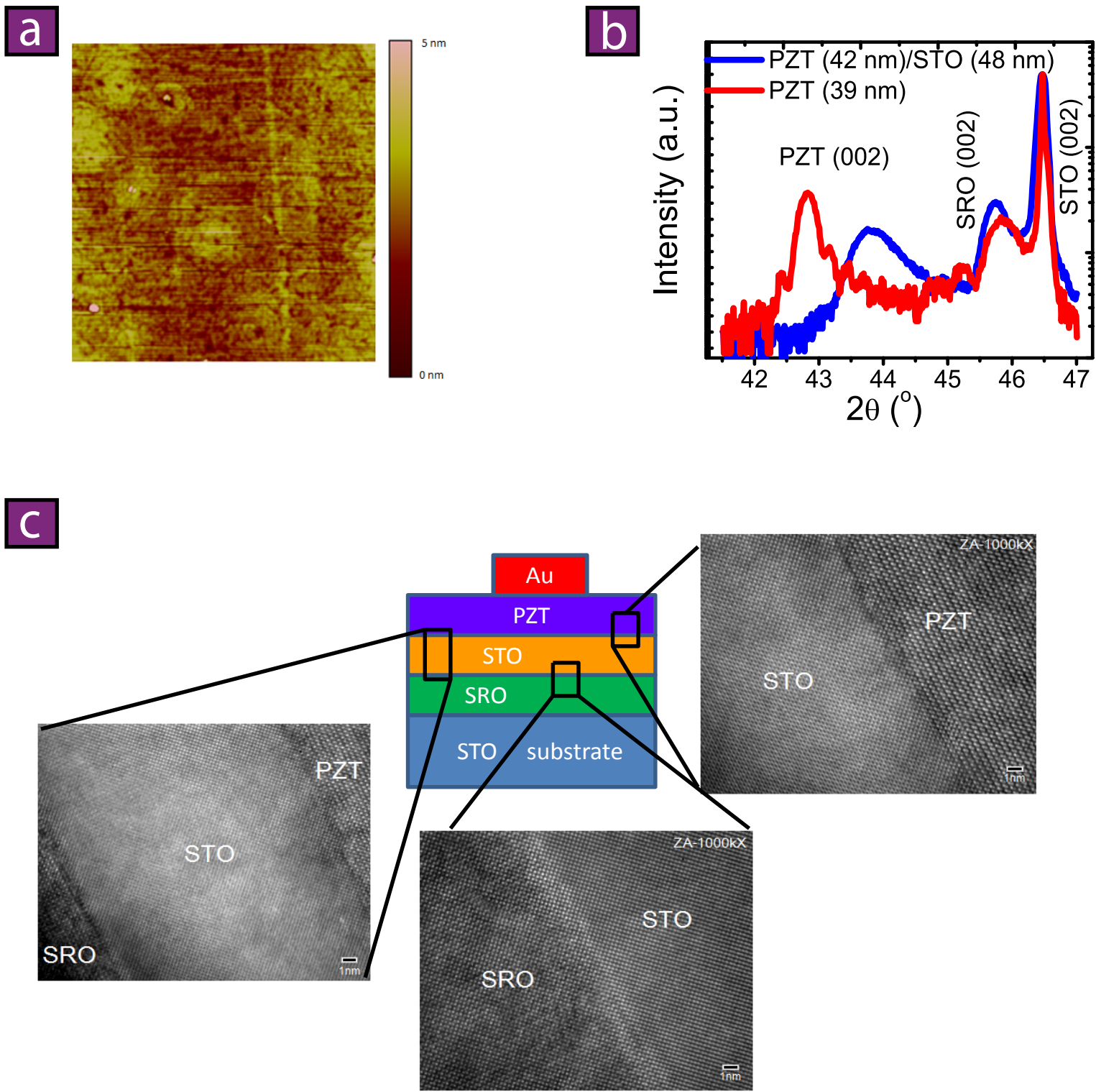

FIGURE S3

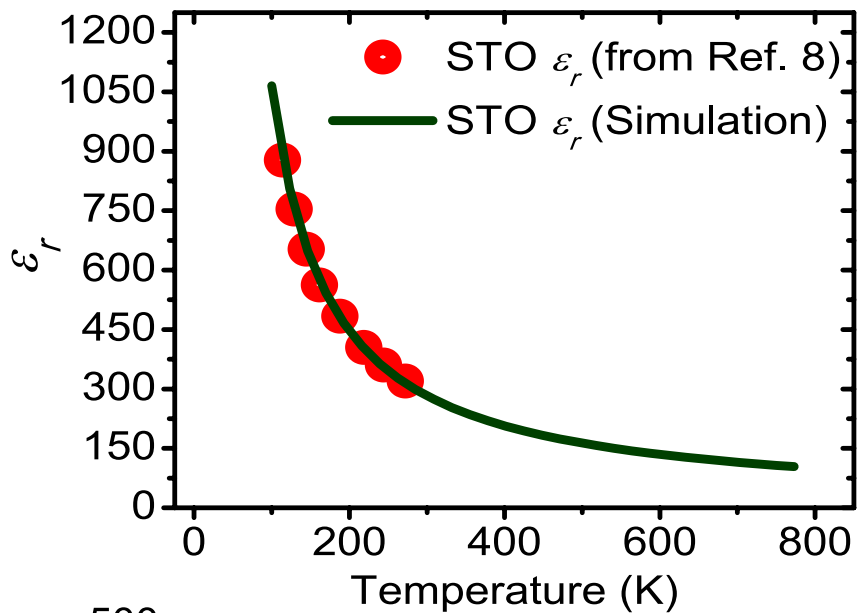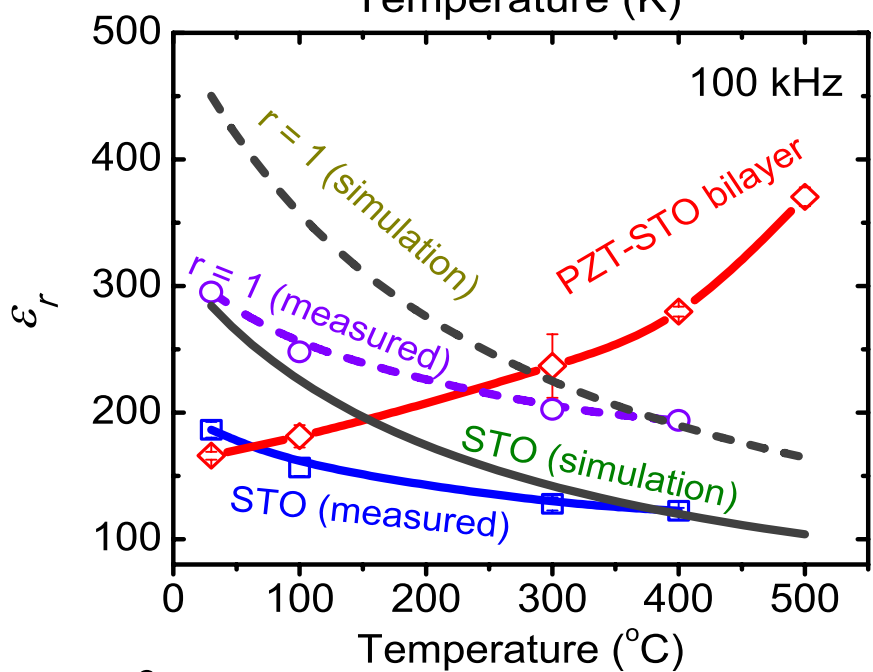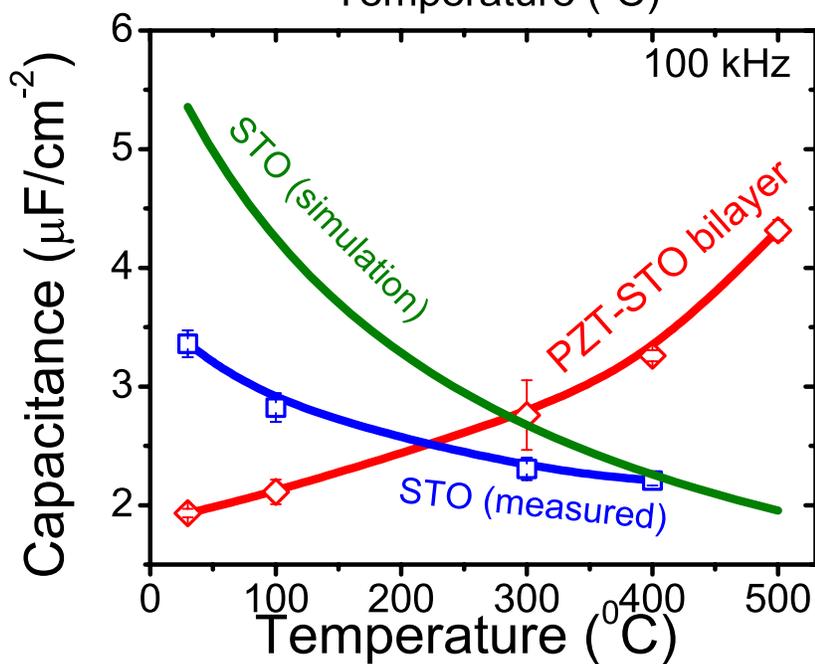

FIGURE S4

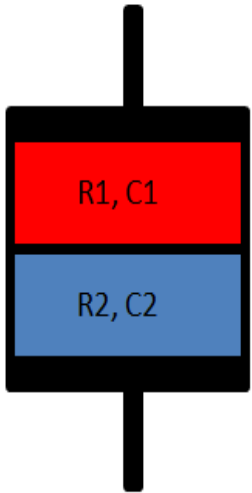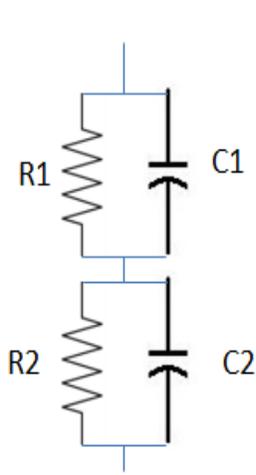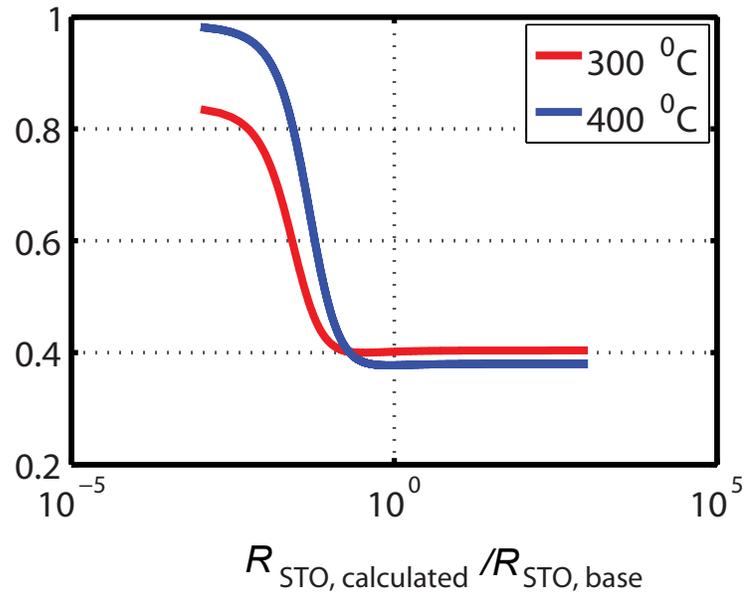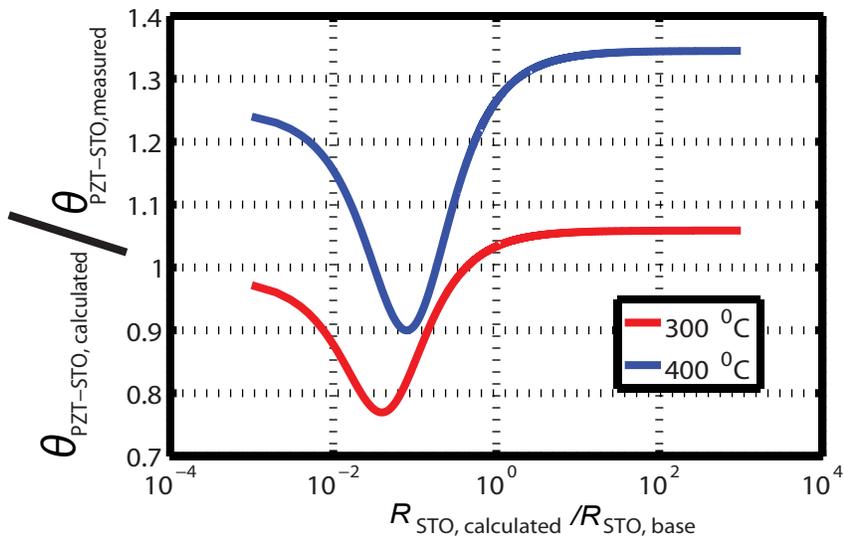

FIGURE S5